\def\etal{{\it et al.~\/}}

\def\ie{{\it i.e.~\/}}

\def\ltsima{$\; \buildrel < \over \sim \;$}
\def\simlt{\lower.5ex\hbox{\ltsima}}
\def\gtsima{$\; \buildrel > \over \sim \;$}
\def\simgt{\lower.5ex\hbox{\gtsima}}

\documentstyle[aasms4,psfig]{article}
\tighten
\singlespace

\lefthead{Ciardi \& Ferrara}
\righthead{Ly$\alpha$ Clouds and Pop~III Objects}

\begin{document}

\title{Ly$\alpha$ Clouds Associated with Pop~III Objects \\
in Cold Dark Matter Models  }

\author{Benedetta Ciardi$^1$, Andrea Ferrara$^2$}
\affil{
$^1$Dipartimento di Astronomia, Universit\`a di Firenze, \\
50125 Firenze, Italy 
\\ E--mail: ciardi@arcetri.astro.it\\
$^2$Osservatorio Astrofisico di Arcetri \\ 50125 Firenze, Italy 
\\ E--mail: ferrara@arcetri.astro.it} 

\begin{abstract}
We present a semi-analytical model to test the hypothesis that
Ly$\alpha$ QSO absorption lines originate in gaseous halos produced by
multiple supernova explosions occurring in Pop~III objects in a CDM cosmological
scenario hoping to assess the validity of CDM models and/or 
constrain their parameters. The preliminary results indicate that
the range of $N_{HI}$, redshift distribution and metallicity of clouds are 
well reproduced by CDM if they are associated with galaxy halos or groups.
Firm conclusions on clouds with $N_{HI}\le 10^{14}$~cm$^{-2}$ need a more 
refined study. 
\end{abstract}

\keywords{Cosmology: theory -- dark matter --quasars: absorption lines}

\section{Introduction}

One of the most challenging current attempts in cosmology is to relate
the origin and evolution of large-scale structure to the observed properties
of QSO absorption-line systems. Several groups have now performed sophisticated 
hydro simulations of various hierarchical models (CDM, CDM+$\Lambda$, HCDM)
to clarify the above issues. In spite of the great advances provided by
numerical approaches, their results are still not conclusive (and often hard even
to compare) since the physics included is necessarily limited by computational 
economy. In particular, the study of effects as feedback and radiative 
transfer, whose importance, particularly for small objects -- such as, presumably, 
the very first generation of galaxies (Pop~III) -- can hardly be overlooked, is 
still in its youth (Gnedin \& Ostriker 1997). These facts motivate the effort 
to develop versatile semi-analytical
models, that could co-operatively complement numerical studies. 
Here we present one such whose aim is to test the hypothesis that  
Ly$\alpha$ QSO absorption lines originate in gaseous halos produced by multiple
supernova explosions occurring in Pop~III objects in a CDM cosmological
scenario.

\section{Dark halos in CDM models}

We assume a CDM power spectrum of fluctuations as in Efstathiou \etal (1992), 
\begin{equation}
\vert \delta_{k} \vert ^{2}= \frac{\displaystyle A k^{n}}{\displaystyle
\left\{1+ \left[Bk + (Ck)^{3/2} + (Dk)^{2} \right]^{\nu}
\right\}^{2/ \nu}},
\label{powerspectrum}
\end{equation}
where $\nu$=1.13; $B$=6.4 $h^{-2}$ Mpc; $C$=3.0 $h^{-2}$ Mpc; $D$=1.7
$h^{-2}$ Mpc; $n$=1. 
The previous spectrum describes a ``standard'' CDM model with $\Omega_0=1$
and $h=0.5$, 
our adopted values; we also take
$\Omega_b=0.19$ (from Coma cluster data, White \& Frenk 1991).
The power spectrum normalization has been determined from {\it COBE},
yielding $\sigma_8=1.07$ (Efstathiou \etal 1992).
To obtain the number density of dark matter (DM) halos, $n$, 
as a function of redshift, $z$, and of
their circular velocity, $v_c=[GM(r)/r]^{1/2}$, we use the Press -- Schechter formalism 
(Press \& Schechter 1974, PS) following White \& Frenk (1991):
\begin{eqnarray}
n(v_{c},z) =-\frac{\displaystyle 3 \delta_{c} (1+z)}{\displaystyle
(2 \pi r_0^2)^{3/2}
\Delta(r_{0})} \frac{\displaystyle d{\rm ln}
\Delta(r_{0})}{\displaystyle d{\rm ln} r_{0}}{\rm exp}\left[-
\frac{\delta_{c}^{2}
(1+z)^{2}}{2\Delta^{2}(r_{0})} \right];
\label{dhaloesdensity}
\end{eqnarray}\\
$\delta_{c}=1.68$ is a critical value of the overdensity and
$\Delta^{2}(r_{0})$ is the variance of fluctuations on scale $r_{0}$:
\begin{equation}
\Delta^{2}(r_{0})=\frac{1}{2\pi^{2}} \int_{0}^{\infty} dk~ k^{2} \vert
\delta_{k} \vert^{2} W^{2}(kr_{0}).
\label{variance}
\end{equation}
We used three different window functions $W(kr_{0})$:
(i) gaussian, (ii) sharp k-space, (iii) top-hat. The results are very weakly
dependent on the specific choice of $W(kr_{0})$; we adopt a
top-hat window function since it gives the best agreement with numerical 
simulations for $\delta_c=1.68$   (Lacey \& Cole 1994). 

\section{Evolution of the baryonic component}

As DM halos are formed, baryons settle dissipatively into their
potential well, being heated by virialization shocks. Both baryonic evolution 
and galaxy formation can be
influenced by another important heating/ionization source, the ultraviolet background
radiation field (UVB); its potential importance has been recently discussed by
Efstathiou (1992) (see also Thoul \& Weinberg 1996), who emphasized its role
in delaying or even suppressing the formation of very low $v_c$ objects. 

\subsection{Reionization epoch and UVB}

In the absence of ionizing radiation, baryons can collapse freely 
in the gravitational potential of the DM, allowing for the 
formation of the first proto-galaxies. These objects, which we will 
refer to collectively as Pop~III, should have masses 
$M_P \simgt 10^{6} M_{\odot}$ in order to be able to cool in a Hubble time 
(Tegmark \etal 1997). According to PS, the
formation peak for these objects
is at $z_f \sim 30$. Assuming an efficiency of
UV photon production per collapsed proton $\sim 190$ (Tegmark \etal 1994),
and an average life time of massive stars $\sim 10^7$~yr, the ionizing 
photon rate per galaxy is $ \sim 7~\times 10^{50}$~s$^{-1}$.
These photons will produce expanding HII regions in the surrounding IGM,
eventually overlapping. 
To determine the epoch of complete reionization we calculate
the IGM porosity parameter, $Q$, defined by
$dQ(z) = \vert dn(M_P,z)/dz\vert r_i(z)^3 dz$,
with $r_i\sim (1+z)^{-3/2}$~Mpc 
being the radius of the HII region. Complete overlap (\ie $Q=0.16$, Smith
1976) occurs at $z_{ion}\sim 7$; this value depends on $M_P$ ($z_{ion}\sim 4.5 
[8]$ for $M_P =5~\times 10^6 [5~10^5] M_{\odot}$).
For $z< z_{ion}$ the UVB   
is approximated as a power law spectrum with index $\alpha$ and
intensity at the Lyman limit (Haardt \& Madau 1996, Ferrara \& Giallongo 1996
[FG]):
\begin{equation}
J_{-21}(z)=0.53 \left\{
\begin{array}{ll}
0 & z>7, \\
{\rm exp}[-0.69(z-3)] & 3<z\leq 7, \\
(1+z)^{1/4} & 2\leq z \leq 3, \\
(1+z)^{3} & z<2.
\end{array}\right.
\label{normflux}
\end{equation}\\
where $J_{-21}=J_0/10^{-21} {\rm ergs\ s^{-1}\ cm^{-2}\ Hz^{-1} sr^{-1}}$.

\subsection{Galaxy formation}

As the UVB heats the IGM, it can prevent the
gas from collapsing and forming proto-galaxies if the IGM temperature, $T$, is
higher than the virial temperature, $T_c \propto v_c^2$, 
of the corresponding DM halo.
We have derived the temperature and ionization evolution of the IGM (assumed
of primordial chemical composition) from the standard time-dependent energy 
and ionization equations. The cooling function is described in
FG; we have studied two cases corresponding to values
of $\alpha=1.5 (5)$ mimicking a hard, AGN-dominated (soft, galaxy-dominated)
spectrum. We have also explored different initial values $T_i=T(z=7)$ finding
that this parameter has essentially no influence on the results, as 
already stressed by FG: we
take therefore $T_i=10^{3}$~K.
The condition $T<T_c$ defines a minimum circular
velocity, $v_{c,min}(z)$, (or equivalently a minimum mass) for collapse, shown 
in Fig. 1. Galaxies with {\it total} mass below $1-4\times 10^7 M_\odot$
are prevented from collapsing after the onset of the UVB; of course
$v_{c,min}=v_{c,min}(M_P) \sim 1.3 (M_P/10^6~M_\odot)^{1/3}$~km~s$^{-1}$ for $z>7$.


\subsection{Blow away and secondary halos}

As the gas collapses towards the center of the proto-galaxy, the density
enhancement is very likely to trigger star formation (SF). Star forming regions 
will inject energy into the galactic ISM, largely in the form of supernova
explosions: for a Miller-Scalo IMF, one supernova is expected for each 
56~$M_\odot = \nu^{-1}$ of stars formed. Type~II SNe tend to occur in associations,
generating very energetic multiple explosions whose observational signature 
is an expanding supershell.  
If the galaxy mass is low, its entire gas content could be swept by the shell,
and driven into the surrounding IGM, in a process called {\it blow-away}. 
The expanding shell will be eventually confined by the IGM pressure, provided
it is high enough. The blow-away condition essentially 
requires that $v_s > {\cal A}(\Omega_0, \Omega_b, g) v_e$, where $g$ is a
geometrical factor (details in Ferrara \& Tolstoy 1997). The previous condition
determines the maximum circular velocity for which the blow-away can
occur, $v_{c,max}=16 (1+z)^{1/2}$~km~s$^{-1}$ for our parameters, also shown 
in Fig. 1;  more massive objects will escape disruption. 
Hence, only objects with $v_{c,min}\le v_c \le v_{c,max}$ will be able
to generate a secondary halo, via a shell {\it re}-expansion, as explained
below.
Since the gas is lost, further SF is inhibited, and the remnant is 
a faint agglomerate of low mass stars produced in the
first (and last) SF episode.  

Next we derive the evolution of the (spherical) supershell
radius, $R_{s}$, in the thin-shell approximation (Kompaneets 1957).
The unperturbed external medium has density
$\rho_{e}(r)\propto r^{-2}$ for $r \le r_{200}$, corresponding to the galaxy
ISM, and $\rho_{e}(r)= \rho(z)$ for $r > r_{200}$, the IGM density.
The shell is driven by the pressure, $P_b$, of the hot gas in the bubble 
cooling at a rate $\Lambda(T_b)$, and subject to the gravitational pull of the
DM halo (we neglect self-gravity of baryons). 
The SN mechanical luminosity is $L_{SN}\propto \gamma= \nu
\dot M_* \sim (\nu M_{gas}/\tau t_{ff})= 1.7~\times 10^{39} (v_c/30~{\rm km~s}^{-1})^3$
ergs~s$^{-1}$, where $\gamma$ is the SN rate, $\tau=0.6 \%$ is the SF efficiency
and $t_{ff}$ is the free-fall time of the gas.
The evolution of $R_{s}$ is governed by energy and momentum conservation 
equations (subscripts $b$ and $s$ refer to bubble and shell,
respectively):
\begin{equation}
\frac{dE_{b}}{dt} = L_{SN} - 3 V_b P_{b} {\dot{R_{s}}\over R_s} -n_{b}^{2} V_b
\Lambda(T_b),
\label{eqen}
\end{equation}
\begin{equation}
\frac{d}{dt} \left(V_b \rho_{e} \dot{R_{s}}
\right) = {3 V_b\over R_s}  (P_{b}-P_{e}) -\frac{G
M(R_{s})}{R_{s}^{2}} V_b \rho_{e},
\label{eqmom}
\end{equation}
where $V_b=(4\pi/3)R_s^3$.
Fig. 2 shows the behavior of $R_{s}$ as a function of $z$ for different
values of $v_c$ and corresponding DM halo formation redshift $z_f$; the evolution is
followed up to the stagnation point, \ie $\dot R_{s}\simeq $ 
sound speed in the IGM. Secondary
halos associated with DM halos formed at $z>7$ are larger, since their
expansion is unimpeded until the reionization epoch due to the low IGM pressure;
the final size increases with $v_c$, since $L_{SN}\propto v_c^3$.
The final value of $R_s$ depends on $P_e$: sizes of the order of tens of kpc
are obtained setting $P_e \sim 10^3 \times$ the IGM pressure, or $P_e/k_B \sim
7$~K~cm$^{-3}$ at $z=0$. This value is typical of the environments of 
spiral galaxies and groups: for the Galactic halo at
$r=30$~kpc, $P_e/k_B \sim 12$~K~cm$^{-3}$ (Wolfire \etal 1995). Lower values
of $P_e$ result in larger sizes and lower densities.

Once the stalling point is reached, 
as the shell pressure is higher than $P_{b}$, the inner boundary of the 
shell re-expands back into 
the cavity. The expansion ends when pressure equilibrium across the contact
discontinuity has been achieved: from this condition we derive the shell 
thickness, $S_s$. We find that the ratio
$S_s/R_s > 1/2$ in all cases and it often exceeds 0.8. In addition,
the re-expansion factor can be enhanced by a magnetic field (Slavin \& Cox
1992) and by accounting for cooling of hot bubble gas. For this reason, 
we assume that the volume inside $R_s$ is uniformly
filled with blown-away baryons with density $n_H=3\Omega_bM/4\pi(1-\Omega_b)R_s^3$.

\subsection{Ly$\alpha$ clouds}

We are now able to test the main hypothesis of this work, \ie
that the secondary baryonic halos so found can be responsible for Ly$\alpha$ 
absorption lines. 
As a preliminary test (a complete study is devoted to a forthcoming paper) 
we have derived the neutral hydrogen column density of the clouds and their 
redshift distribution.
Assuming a value for the secondary halo gas (or the Ly$\alpha$ clouds, in our view)
temperature of $\sim 4\times 10^{4}$~K, consistent with a purely thermal Doppler parameter 
$b \sim 25$~km~s$^{-1}$ (Lu {\it et al.} 1996), we have determined the HI
density, $n_{HI}$,
from ionization equilibrium, Integrating along a line of sight at
impact parameter $p$, we find hydrogen column densities $10^{12}< N_{HI}<  
5\times 10^{18} {\rm cm^{-2}}$, as shown
in Fig. 3, with the highest $N_{HI}$ originating in low $z_f$ halos.
Column densities $\le 10^{14}$~cm$^{-2}$ are associated with secondary halos 
formed at very high redshift, $z_f\sim 20$, where merging -- that we have
neglected -- should be rather important. 
The redshift distribution of the clouds, 
\begin{eqnarray}
\frac{dN}{dz}(z,N_{HI}) = \frac{c}{H_{0}} (1+z)^{1/2}
\int^{v_{c,max}}_{v_{c,min}} dv_c~n(v_{c},z)v_{c}^{-1}
\pi p^{2}(v_{c},N_{HI}),
\end{eqnarray}
can be very nicely fit by a power law $dN/dz = A_0
(1+z)^\gamma$ with $\gamma=2.2$ and $A_0$=0.7, in satisfactory agreement with
observational data (Bechtold 1994, Giallongo \etal 1996) at this stage.

\section{Conclusions and discussion}

We have brought arguments suggesting that at least some Ly$\alpha$ 
QSO absorption lines could originate in secondary halos produced by multiple 
supernova explosions occurring in Pop~III objects. As a first step, we have
investigated only a specific CDM model; nevertheless our aim is to extend
the analysis to a larger class of theoretical models (CDM+$\Lambda$, HCDM)
In this paper we have introduced the main features of our semi-analytical model,
deferring a complete exploration of the results, cosmological models, 
and model details to a forthcoming paper. Yet, two   
conclusions could be relevant: 
(i) Ly$\alpha$ clouds should be preferentially found in relatively 
high pressure regions, as galaxy halos and groups; (ii) HI column densities 
and redshift distribution are well reproduced by a CDM model 
if clouds are associated with galaxy halos or groups. 
We reiterate that predictions for $N_{HI}\simlt 10^{14}$~cm$^{-2}$ 
might be affected by the simplifications 
adopted (\ie no merging).
In view of the exploration of additional cosmological models, it is important
to compare the model predictions on other observables as the $dN/dN_{HI}$
and $b$ distributions, and clustering properties. 
The metallicity $Z/Z_\odot\simlt 10^{-2} $ found in clouds with
$N_{HI}>3\times 10^{14}$~cm$^{-2}$ by Cowie \etal (1995) and Tytler \etal (1995)
could be explained by primordial enrichment by 
Pop~III objects.
A back-on-the envelope calculation using the SN rates from our model and
assuming a metal yield of $\sim 1 M_\odot$ per SN, gives $Z/Z_\odot \sim 1.2
\times 10^{-4} (v_{c}/10 {\rm ~km \; s^{-1}})^{2} \;$ in the
secondary halo.


\newpage

\begin{figure}
\caption{\label{fig1} Minimum circular velocity (solid lines) of a DM halo for 
collapse of the gas as a function of redshift for two values of the UVB power-law
spectral index $\alpha=1.5,5$. Also shown (dashed line) is the maximum circular velocity
required for blow-away to occur. The initial IGM temperature is $T_i=10^3$~K.}
\end{figure}

\begin{figure}
\caption{\label{fig2}Behavior of shell radius, $R_{s}$, as a function
of $z$ for different values of $v_c$
(from top to bottom) and corresponding DM halo formation redshift $z_f$;
the evolution is followed up to the stagnation point.}
\end{figure}

\begin{figure}
\caption{\label{fig3}$N_{HI}$ of Ly$\alpha$ clouds as a function of $v_c$
for different values of the DM halo formation redshift $z_f$.  }
\end{figure}
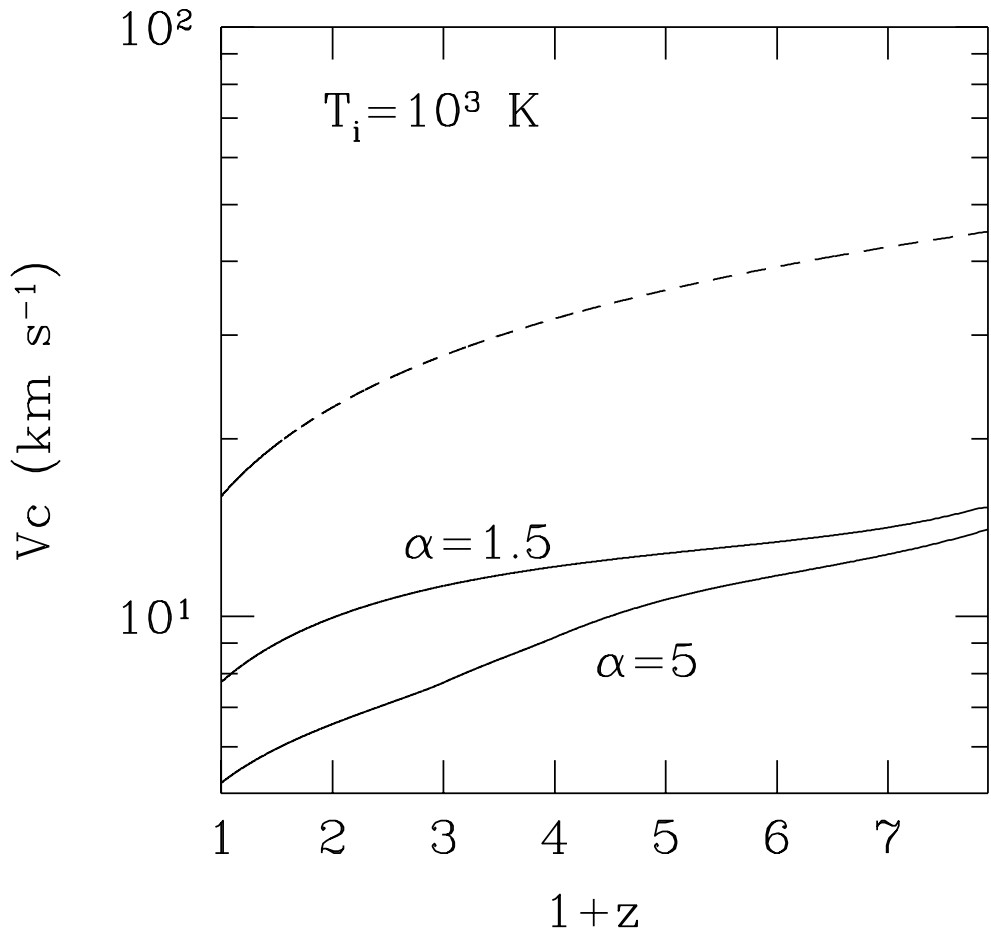
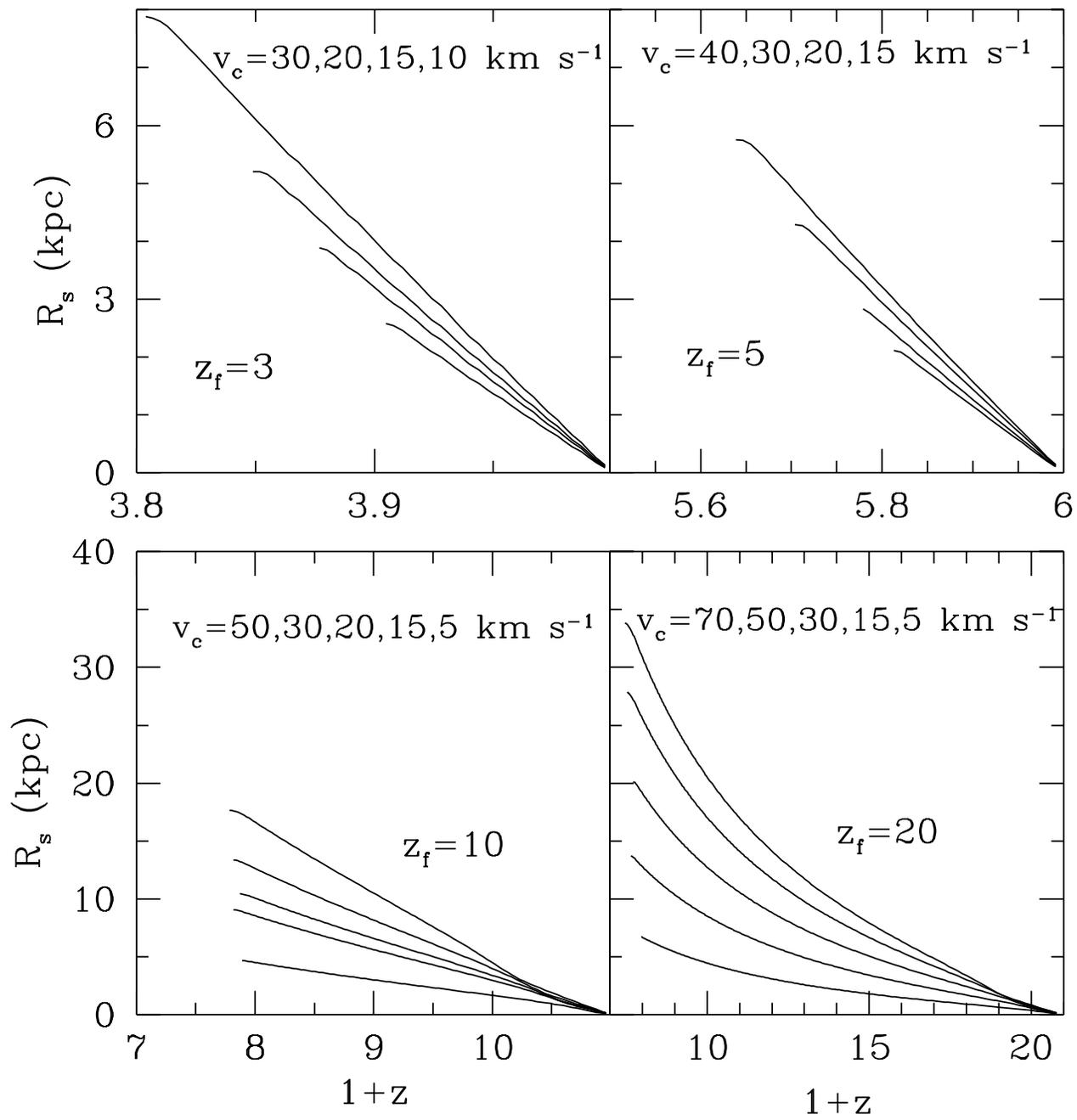
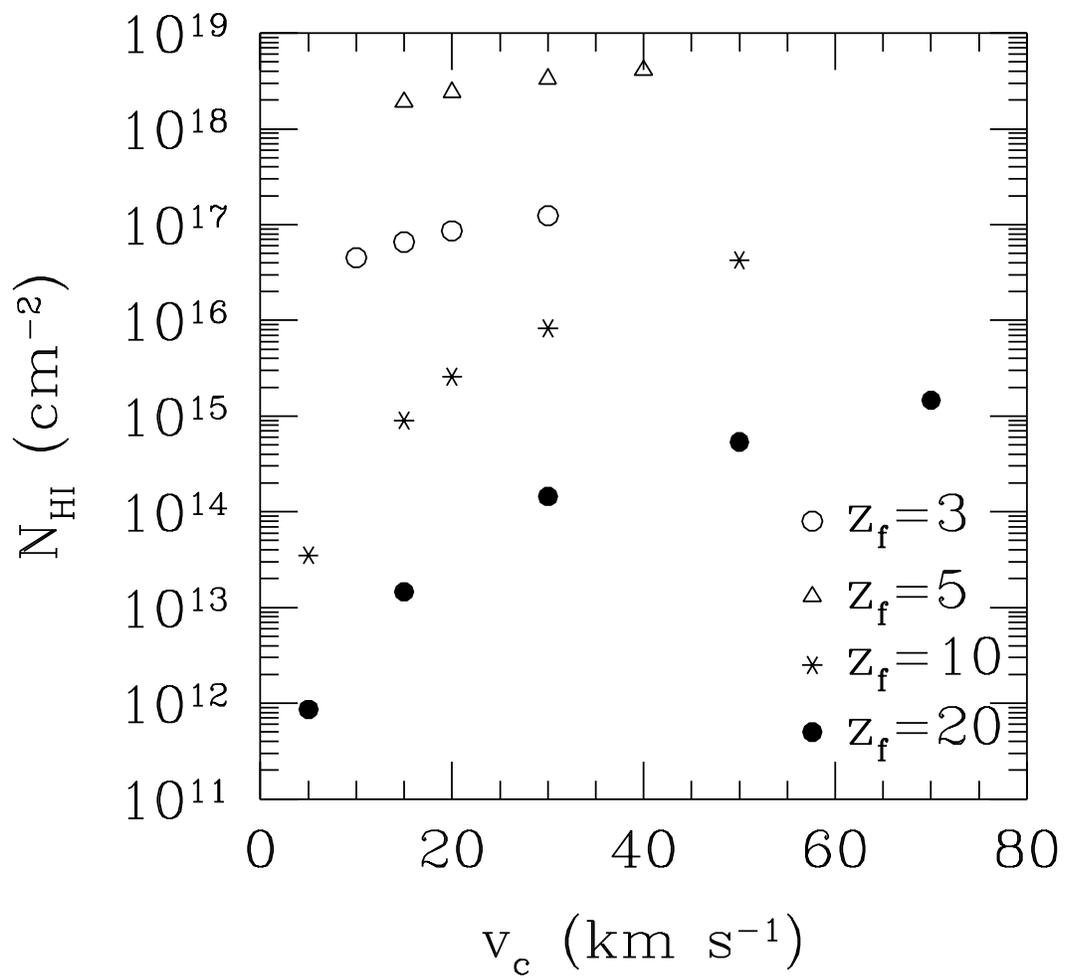
\end{document}